\documentclass[prd,nofootinbib,preprint]{emulateapj}
\usepackage{graphicx}
\usepackage{rotating}
\usepackage{natbib}
\usepackage{amsmath}
\usepackage{subfigure}
\usepackage{epstopdf}
\usepackage{multirow}
\usepackage{setspace}

\def\be{\begin{equation}}
\def\ee{\end{equation}}
\def\ba{\begin{eqnarray}}
\def\ea{\end{eqnarray}}

\def\msun{M$_\odot$ }

\begin{document}

\title{A Close Companion Search around L Dwarfs using Aperture Masking Interferometry and Palomar Laser Guide Star Adaptive Optics}

\author{David Bernat$^{1}$}
\author{Antonin H. Bouchez$^{2,8}$}
\author{Michael Ireland$^{3}$}
\author{Peter Tuthill$^{3}$}
\author{Frantz Martinache$^{4}$}
\author{John Angione$^{5,8}$}
\author{Rick S. Burruss$^{5,8}$}
\author{John L. Cromer$^{2,8}$}
\author{Richard G. Dekany$^{2,8}$}
\author{Stephen R. Guiwits$^{5,8}$}
\author{John R. Henning$^{6,8}$}
\author{Jeff Hickey$^{6,8}$}
\author{Edward Kibblewhite$^{7,8}$}
\author{Daniel L. McKenna$^{6,8}$}
\author{Anna M. Moore$^{2,8}$}
\author{Harold L. Petrie$^{6,8}$}
\author{Jennifer Roberts$^{5,8}$}
\author{J. Chris Shelton$^{5,8}$}
\author{Robert P. Thicksten$^{6,8}$}
\author{Thang Trinh$^{5,8}$}
\author{Renu Tripathi$^{6,8}$}
\author{Mitchell Troy$^{5,8}$}
\author{Tuan Truong$^{5,8}$}
\author{Viswa Velur $^{2,8}$}
\author{James P. Lloyd$^{1}$}
\affil{$^{1}$Department of Astronomy, Cornell University, Ithaca, NY 14853, USA}
\affil{$^{2}$California Institute of Technology, Pasadena, CA, 91125}
\affil{$^{3}$Sydney Institute for Astronomy, School of Physics, University of Sydney, Australia}
\affil{$^{4}$National Astronomical Observatory of Japan, Subaru Telescope, Hilo, HI, 96720}
\affil{$^{5}$Jet Propulsion Laboratory, California Institute of Technology, Pasadena, CA, 91109}
\affil{$^{6}$Palomar Observatory, California Institute of Technology, Palomar Mountain, CA, 92060}
\affil{$^{7}$University of Chicago, Chicago, IL, 60637}
\affil{$^{8}$Palomar Laser Guide Star Adaptive Optics Team, Palomar Observatory, California Institute of Technology, Palomar Mountain, CA, 92060}

\begin{abstract}
We present a close companion search around sixteen known early-L dwarfs using aperture masking interferometry with Palomar laser guide star adaptive optics.  The use of aperture masking allows the detection of close binaries, corresponding to projected physical separations of 0.6-10.0 AU for the targets of our survey.  This survey achieved median contrast limits of $\Delta$K $\sim$ 2.3 for separations between 1.2 - 4 $\lambda$/D, and $\Delta$K $\sim$ 1.4 at $\frac{2}{3}\lambda /D$.  

We present four candidate binaries detected with moderate to high confidence (90-98\%).  Two have projected physical separations less than 1.5 AU.  This may indicate that tight-separation binaries contribute more significantly to the binary fraction than currently assumed, consistent with spectroscopic and photometric overluminosity studies.

Ten targets of this survey have previously been observed with the Hubble Space Telescope as part of companion searches.  We use the increased resolution of aperture masking to search for close or dim companions that would be obscured by full aperture imaging, finding two candidate binaries.

This survey is the first application of aperture masking with laser guide star adaptive optics at Palomar.  Several new techniques for the analysis of aperture masking data in the low signal to noise regime are explored.
\end{abstract}

\maketitle

\section{Introduction}

The mass determinations of stars through binary studies have provided numerous mass-luminosity benchmarks for the testing and calibration of stellar models.  Such studies have only recently begun for the regime of brown dwarfs.

The empirical calibration of brown dwarf models is generally made more difficult by the added dependency on age in the mass-luminosity relationship.  For example, an object spectroscopically classified as a late-M dwarf may be a young brown dwarf or an old star just above the hydrogen burning limit.  This broadens the range of potential physical properties (mass, age, composition) that generate the same observable spectrum.  Conversely, photometry can only very generally reveal the objects' physical properties.  Measurements of brown dwarf masses through the tracking of binary orbits provide the strongest constraints on stellar models, "mass benchmarks" that reduce the degeneracy of photometric studies even for targets with unknown ages \citep{Liu:2008}.    Mass measurements of brown dwarfs by \citet{Konopacky:2010} show systematic discrepancies between models and measurements; late-M through mid-L systems tended to be more massive than models predict, while one T dwarf system was less massive than its model prediction.  This collection of mass benchmarks grows even more important as brown dwarf models are extended to infer the masses of directly imaged planets, such system HR 8799 \citep{Marois:2008}.  

Binary surveys have also begun to turn up interesting statistical results that may suggest the brown dwarf binary formation mechanism is different than that for solar type binaries (see \citet{Burgasser:2006} for a summary of results from low mass surveys, including many results presented in this section).  Few surveys, however, have produced results for very low mass binaries, especially those with very tight separations ($\lesssim$ 3.0 AU).  This regime of short period binaries is particularly challenging to achieve with ground-based direct imaging.

While the companion fraction of brown dwarfs is proposed to be low ($\approx$ 15\%) and peaked within a narrow separation range, 3-10 AU \citep{Burgasser:2008}, little conclusive results are known for separations less than 3 AU despite preliminary evidence that many additional companions are likely to reside at very close distances (\citet{Jeffries:2005,Pinfield:2003,Chappelle:2005}).  

Over 90\% of known very low mass binaries have less than 20 AU \citep{Burgasser:2006}.  Competing theories of stellar formation aim to explain the observed companion statistics of brown dwarfs.  As a general trend, stars appear to have a binary fraction that decreases with mass.  This can partly be explained by the decreased binding energy of lower mass primaries, and thus a maximum binary separation that decreases as a function of total mass (\citet{Reid:2000,Close:2003}).  For very low mass stars, the companion fraction peaks near 3-10 AU, and exhibits a significant (and statistically significant) drop at separations beyond 20 AU separation.  Slightly more than half of known very low mass binaries lay within this narrow separation range \citep{Burgasser:2006}.  On the near side of this peak, the data collected is very likely incomplete, where the necessary resolution (300 mas) stretches the limitations of HST/NICMOS and ground-telescopes with AO alone.  What data has been collected suggests direct imaging may have missed companions at very close separations.

Spectroscopic, spectral morphology, and Laser Guide Star AO surveys suggest that very tight binaries within 3 AU may be as plentiful as binaries of moderate separation.  \citet{Burgasser:2007} has used spectral features of unresolved sources to indicate composite spectra, implying multiplicity.  This technique has suggested numerous early/mid-L dwarfs with potential mid-T dwarf secondaries and systems of equal-mass L/T transition objects.  \citet{Jeffries:2005} used sparse radial velocity data-sets of very low mass systems to predict an additional 17-30\% binaries at separations less than 2.6 AU.  Photometric overluminosity studies by \citet{Pinfield:2003} and \citet{Chappelle:2005} have also hinted at surprisingly larger binary fractions (up to 50\%) in the Pleiades and Praesepe, though concerns over membership contamination and the influence of variability limit the conclusiveness of the results.  In each study, with the exception of the Burgasser mid-L/mid-T systems, very low mass binaries tend towards equal mass pairs ($q\sim$ 1) at close separations, just as is the case at moderate separation.  

These preliminary results contrast those of previous, observationally complete surveys that focused on moderate and wide separation binaries. Those surveys predict that fewer than 3\% of very low mass companions sit at separations closer than 3 AU \citep{Allen:2007}.  This discrepancy speaks to the importance of additional, observationally complete surveys searching for binaries at close separations.

Non-Redundant Aperture Masking (NRM) on 5-10m class telescopes, combined with LGS AO, allows sub-diffraction limit resolution observations at contrasts high enough to search for most binaries in this potentially fertile, unresolved region.  The detection of close brown dwarf binaries, with a typical period of 1-2 years, also allows mass measurements of late-L or T dwarfs, providing particularly valuable empirical benchmarks for the study of low mass stellar models.  To put into perspective the dearth of benchmarks, the mass measurements of fifteen very low mass systems (including six with L or T dwarf components) using LGS AO alone by \citet{Konopacky:2010} has tripled the number of very low mass systems with mass measurements.

In Section~\ref{Obs} we describe the sixteen field L-dwarf targets imaged at Palomar using aperture masking with laser guide star adaptive optics and outline the data analysis techniques used to determine the binarity of the targets.  In Section~\ref{results}, we present the results of our survey, which operated in the range of 60-320 mas (1.1-8.4 AU @ 18.4 pc, the median distance of our targets).  We identify four new candidate L dwarf-brown dwarf binaries at moderate or high (90-98\%) confidence.  This survey achieved median contrast limits of $\Delta$K$\sim$2.3 between 1.2 $\lambda/D$ and 4 $\lambda/D$, ruling out companions down to approximately .06 \msun for old (5 Gyr) systems and .03 \msun for young (1 Gyr) systems.  In Section~\ref{discussion}, we discuss the aperture masking techniques employed in this paper and present recommendations for future faint target observations.  In Section~\ref{conclusion}, we summarize the results of this survey and discuss its implications for future companion searches around brown dwarfs.  Finally, we discuss in detail our methods for calculating detection confidence and contrast limits in the Appendix.
 
\section{Observations and Data Analysis}
\label{Obs}

\subsection*{Observations}

We observed our target sample of sixteen field L dwarfs in September and October 2008 with the Palomar Hale 200" telescope (refer to Table \ref{targetlist}).  

Ten of the sixteen targets in this survey have been observed previously as part of various companion searches using the Hubble Space Telescope \citep{Reid:2006, Bouy:2003}.  These previous observations were capable of resolving low contrast or distant (beyond about 300 mas) companions.  Aperture masking complements these previous surveys, extending the detection limits around these targets to dimmer and closer companions.

\begin{figure}[b]
\includegraphics[width=0.45\textwidth]{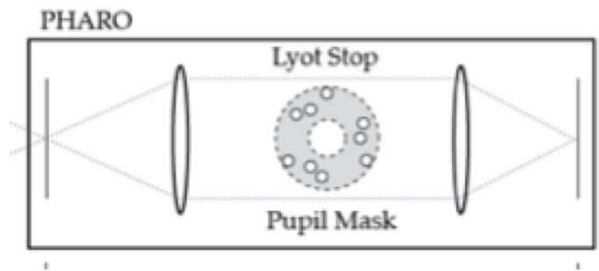}
\caption{The aperture mask inserted at the Lyot Stop in the PHARO detector.  Insertion of the mask at this location is equivalent to masking the primary mirror.\label{pharo_mask}} 
\end{figure}

Aperture masking observations were obtained using the PHARO instrument \citep{Hayward:2001}, with a 9-hole aperture mask installed in the pupil plane of the Lyot-stop wheel (Figure \ref{pharo_mask}).  The longest and shortest baselines, which set the approximate inner and outer working angle, are 3.94m and 0.71m respectively (58 and 320 mas in K band).  We operate to minimize atmospheric and AO variation during a single image, using PHARO in 256 x 256 sub-array mode with a total of 16 reads (sub-frames) per array reset and 431 ms exposures.  Every read was saved to disk.  In post-processing, we discard the first three sub-frames of each exposure (usually corrupted by detector reset), and combine the remaining sub-frames by a Fowler sampling algorithm in which later sub-frames are weighed more heavily.  Approximately 300 images (each with 16 sub-reads) were taken in K$_s$ for each target, for a total integration time of roughly 60-70 minutes per target.

The Palomar laser guide star adaptive optics system \citep{Roberts:2008} provided the wavefront reference for high order AO correction, while nearby (a few arc-minutes) field stars were observed contemporaneous to provide tip-tilt correction.

Aperture masking operates most effectively when exposure times are as short as possible, but long enough to observe fringes over read noise.  The optimal exposure time depends on the brightness of the targets and the level of correction provided by the AO system.  Poor correction favors shorter exposure times, where variation of the incoming wavefront quickly degrades the average transmission of long baseline frequencies.  For targets brighter than about tenth magnitude, the read out limited exposure time of the PHARO detector, 431 ms for the 256 x 256 array, is sufficient to observe long baseline fringes.  The targets of this survey are approximately twelfth magnitude, and initial experimentation showed that short exposures did not consistently provide long baseline fringes.  Longer exposure times (1 minute) faired poorly because variations in correction over the exposure degraded the average transmission of long baselines below background levels.  We opted to use short exposures and to weigh more heavily in post-analysis those observations in which long baseline fringes could be seen (see additional discussion later in this section).  

Background subtraction is necessary for targets as faint as L Dwarfs and background levels were often comparable to the signal levels.  In many instances our observations were background limited.  To remove the background in post-processing, each target was dithered on the 256 x 256 sub-array.  

A requirement for obtaining good contrast limits around bright targets is the contemporaneous observation of calibrator sources: single stars which are nearby in the sky and similar in near-infrared magnitudes and colors.  This calibration is necessary to remove non-stochastic phase errors introduced by primary mirror imperfections and other non-equal path length errors.  This error can be on the order of one to a few degrees, comparable to the measurement scatter of the closure phases for bright targets.  For brighter targets, the typical observing mode is to obtain several observations of the science target, interspersed with observations of calibrator stars.  However, the lengthy time of acquisition for the laser guide star AO system made this method inefficient for this survey.  Furthermore, the measurement scatter for the faint targets of this survey were much larger than the expected systematic error.  Therefore, we did not use calibrator stars.  We note that calibrator stars have also not been used for similar reasons in \citet{Dupuy:2009}.

\subsection*{Aperture Masking Analysis and Detection Limits}
\subsubsection*{Extracting Closure Phases from Raw Images}

The core aperture masking pipeline implemented in this paper is similar to that discussed in previous work \citep{Lloyd:2006, Pravdo:2006, Kraus:2008}, with additions to handle low signal to noise data and calculate confidence intervals and contrast limits.

Raw images are first dark subtracted and flat-fielded, bad pixels are removed, and the data is windowed by a super-Gaussian (a function of the form $\exp(-kx^4)$).  This window limits sensitivity to read noise and acts as a spatial filter.  A per-pixel sky background map is then constructed from the set of target data and subtracted.  The background map is generated by masking out the target from each image within a set, then, for each pixel, using the median value of the pixel flux from those images that were not masked.

The point spread function of the nine hole mask consists of thirty-six interfering fringes, called the {\em interferogram}.  Because the mask is non-redundant, each fringe is produced uniquely by the pairing of two holes; the amplitude and phase of this fringe translates directly to the complex visibility of the corresponding spatial frequency.

\begin{figure}[t]
\includegraphics[width=0.45\textwidth]{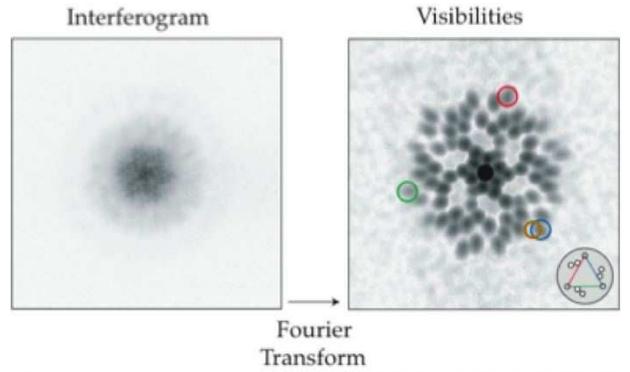}
\caption{Interferogram and power spectrum generated by the aperture mask.  (Left) The interferogram image is comprised of thirty-six overlapping fringes, one from each pair of holes in the aperture mask.  (Right) The Fourier transform of the image shows the thirty-six (positive and negative) transmitted frequencies.  (Right, inset and overlay) Closure phases are built by adding the phases of 'closure triangles': sets of three baseline vectors that form a closed triangle.\label{closure_triangles}} 
\end{figure}

Fourier-transforming each image reveals seventy-two patches of transmitted power we call {\em splodges} (thirty-six frequencies transmitted, positive and negative)(Figure \ref{closure_triangles}).  The complex visibilities are extracted by weighted averaging of the central nine pixels of each splodge.  Optical telescope aberrations, AO residuals, and detector read-noise contribute noise to the complex visibilities.  Under the best conditions, visibility amplitudes suffer large ($>$ 100\%) calibration errors and are not used for the analysis in this survey.

Visibility phase suffers less from these variations, but the use of the complex triple product and {\em closure phase} \citep{Lohmann:1983} yields an observable that reduces the effect of wavefront-degradations from baseline-length independent sources such as low-order AO residuals.  For an interferometric array (or aperture mask), closure phases are built by adding the visibility phases of 'closure triangles': sets of three baseline vectors that form a closed triangle (see Figure \ref{closure_triangles}).  The set of closure phases have lower noise than visibility phases, allowing precise photometric measurements despite the loss of photons imposed by the mask.

Thirty-six baselines are present with the 9-hole mask, from which 84 closure phases can be constructed.  However, these closure phases are not all linearly independent, and the 36 baseline phases cannot be uniquely determined.  The phase information cannot be uniquely inverted (by inverse Fourier transform) into an image without further assumptions (see, for example, the CLEAN algorithm \citep{Hogbom:1974} and the Maximum Entropy Method \citep{Gull:1984}).  

As our survey is a search for binaries, the closure phase signal of such a target can be modeled easily.  Thus, instead of inverting the closure phases to form an image, we search for the modeled binary configuration that best fits the measured closure phases.  

\subsubsection*{Typical Results on Bright Targets}

	Aperture masking with natural guide star adaptive optics has been employed during numerous near infrared surveys on the Palomar and Keck telescopes.  (For mass determinations made through orbit tracking see \citet{Lloyd:2006, Ireland:2008, Martinache:2006, Martinache:2008} and \citet{Dupuy:2009}, and \citet{Kraus:2008} for an extensive survey of Upper Scorpius.)  Aperture masking has also recently begun usage in conjunction with Keck laser guide star adaptive optics \citep{Dupuy:2009}).
	
	The observation of bright targets (K$_s \lesssim$ 9), such as nearby early-M dwarfs, with an aperture mask enables the detection of companions of contrast up to 150:1 ($\Delta$K$_s \sim 5.5$) at the formal diffraction limit and 20:1 ($\Delta$K$_s \sim 3.3$)  at $\frac{2}{3}\lambda/D$ in at Palomar.
	
	In this regime, non-stochastic phase errors introduced by the optical pipeline dominate closure phase errors, as well as background flux, wavefront residuals of the adaptive optics system, and achromatic smearing of the fringes.  Typically, these sources contribute errors on the order of one degree after calibration.

\subsubsection*{Noise Properties of Dim Targets}

For each star image, of which we have approximately 300 for each star, we extract closure phases.

\begin{figure*}[ht]
\centering
\subfigure[Closure Phase 50]{\includegraphics[width=0.32\textwidth]{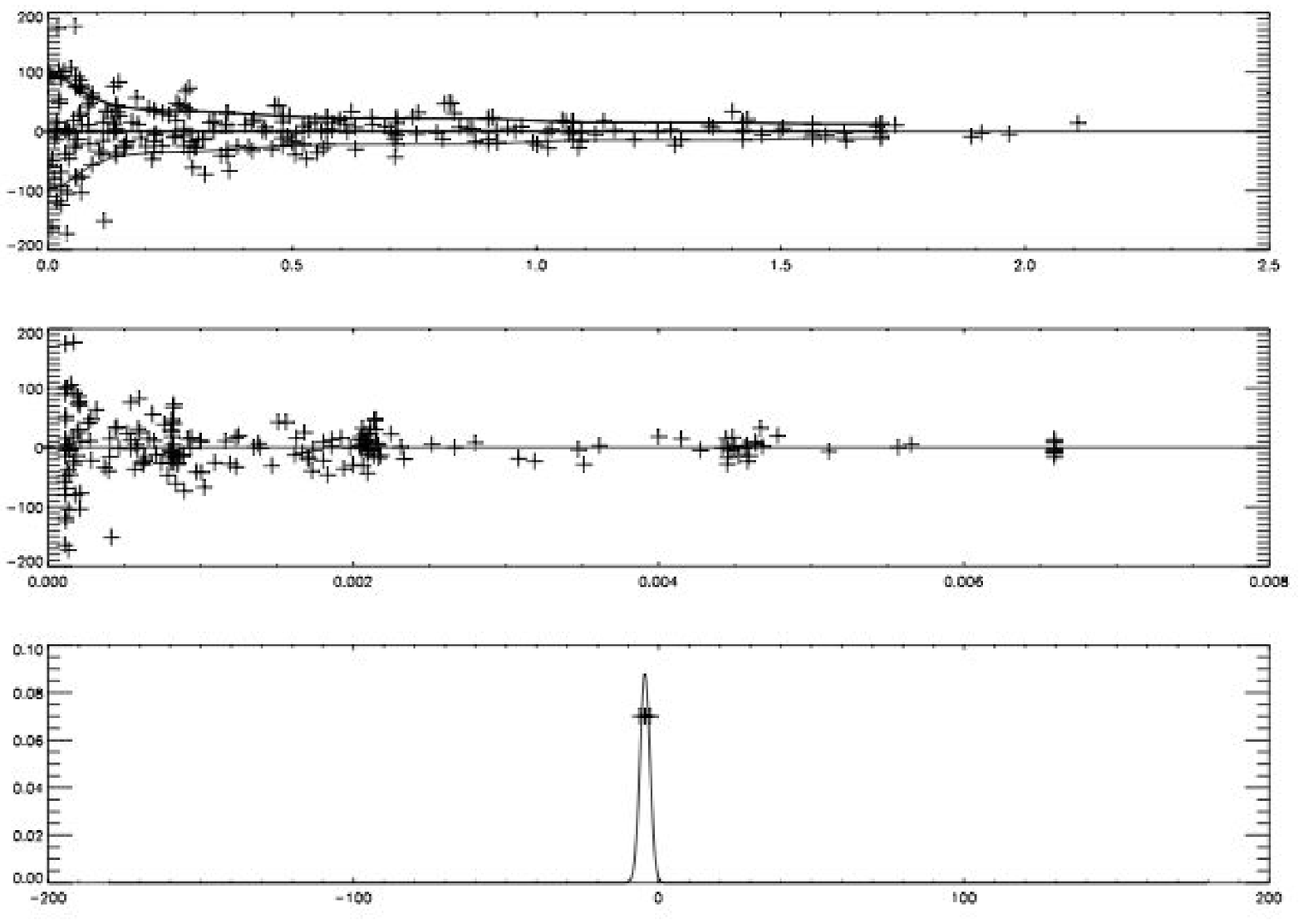}}
\subfigure[Closure Phase 84]{\includegraphics[width=0.32\textwidth]{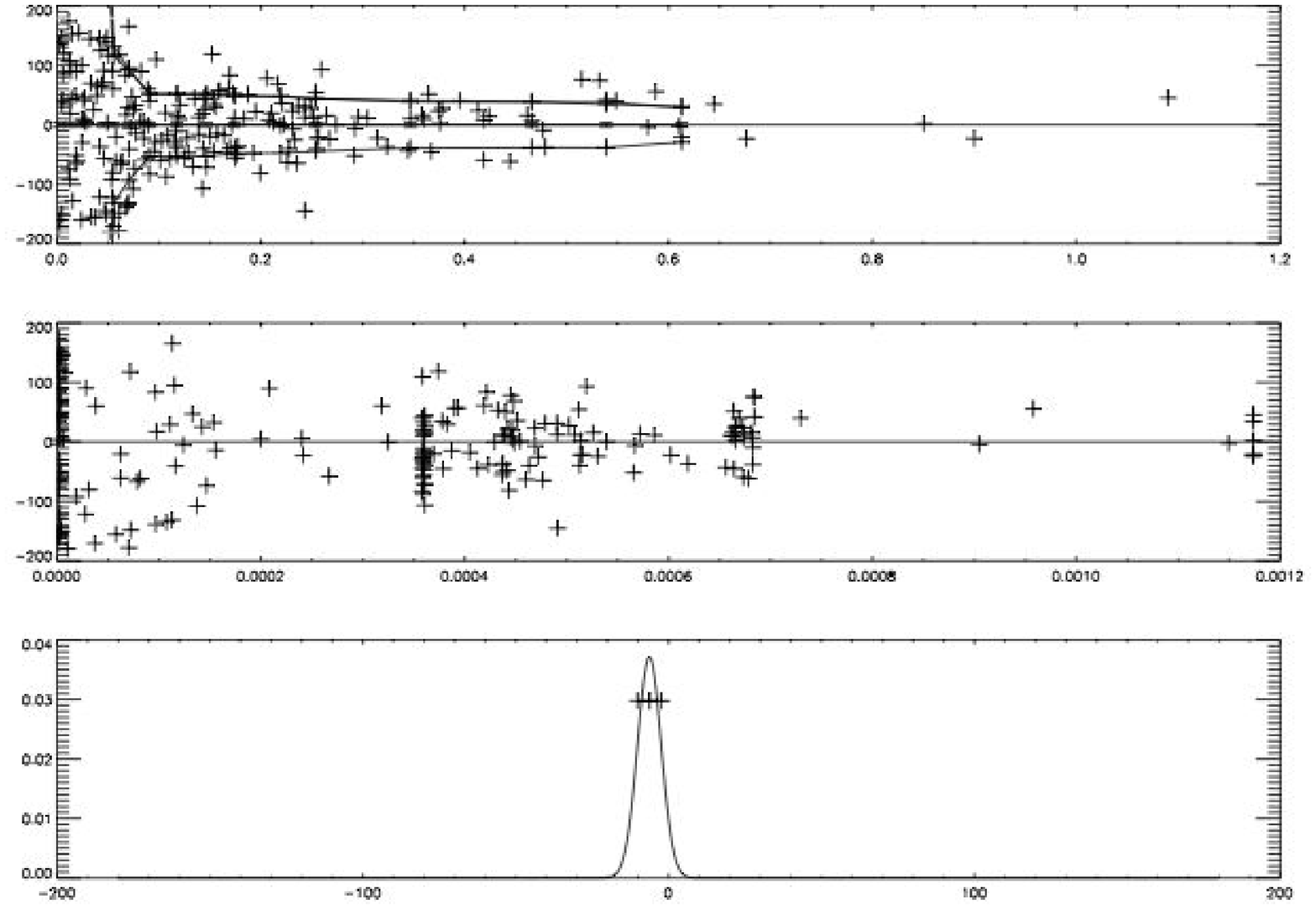}}
\subfigure[Closure Phase 43]{\includegraphics[width=0.32\textwidth]{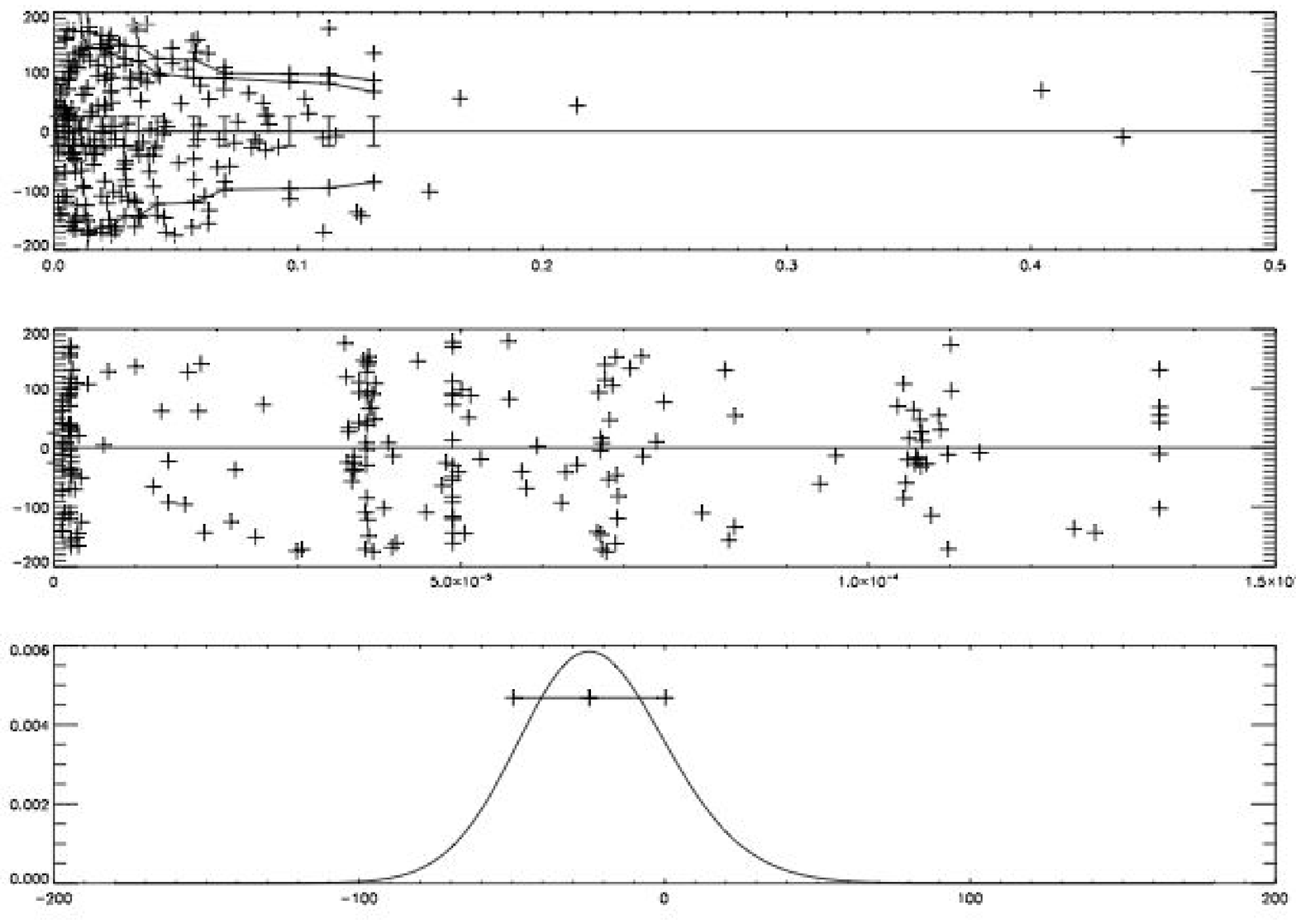}}
\caption{Estimating per-measurement weights for three closure phase data sets for target 2M 2238+4353.  The data sets have comparatively high- (left), moderate- (center) and very low- (right) signal to noises.  (Top) Plot of bispectrum (closure) phase vs. bispectrum amplitude.  Note that larger amplitude data have smaller phase spreads, and a clear asymptotic mean can be identified in the high and moderate signal to noise cases.  (Closure phase 43 contains no discernible signal, and would be removed from further analysis.)  Low amplitude bispectra are swamped by read noise, introducing phase errors which are nearly uniformaly distributed.  The solid line estimates the relationship between per-measurement standard deviation and bispectrum amplitude.  (Middle) Closure phase vs. approximate weighting.  Note that the higher weighted points have lower per-measurement standard deviation.  (Bottom) Resulting p.d.f. of the closure phase. \label{meanfinder} } 
\end{figure*}

Because our targets are faint and exposure times are short, detector read-outs contribute significant noise in the phase and amplitude of the complex visibilities and bispectrum.  Amplitudes are particularly susceptible to calibration errors.  Even during high signal to noise conditions, amplitudes have been seen to fluctuate by up to 100\%, and are not directly used for fitting to model binaries.  However, closure phase data show a clear improvement in per-measurement signal to noise for increasing amplitude.  That is, bispectrum with the largest amplitude tend to have the highest fidelity closure phases.  In order to pare off bad data and weigh higher signal to noise measurements more heavily, we empirically estimate the relationship between amplitude and closure phase fidelity (Figure \ref{meanfinder}).

This relationship is estimated by binning the set of closure phase data by amplitude and calculating the standard deviation of each bin.  As already described, as the average amplitude within a bin increased, the standard deviation within the bin decreased.  To first approximation, this estimates the relationship between amplitude and closure phase fidelity.

The noisiest bins often show closure phase errors approaching 180$^\circ$.  Because the closure phase is inherently a measurement of the bispectrum phasor, there is a 360$^\circ$ ambiguity in the measurement of closure phase.  Furthermore, even if the underlying noise source is Gaussian distributed, the distribution of measured closure phases approaches a uniform distribution when the standard deviation of the noise source is larger than about 180$^\circ$.  Direct calculation of the root mean squared deviation under represents the variance of the underlying noise source; the calculation of the mean depends on the choice of angle zero-point.  The variation within one bin was at times large enough to motivate alternative methods for averaging bispectrum data.  

We adopt a maximum likelihood method to calculate the standard deviations of bins and overall closure phase mean.  We presume the closure phases in each bin are drawn from a wrapped normal distribution\footnotemark.  The standard deviation is varied to maximize the likelihood of the data in the bin.  The same mean is used for every bin, and the mean which maximizes the likelihood of the entire data set is data set's overall mean.  This allows bins to take arbitrarily large standard deviations; a wrapped normal distribution with large standard deviation converges towards a uniform distribution.  For bins dominated by read-noise or very low signal to noise, this method accurately estimates very large standard deviations and translate that into very low weighting for the bin.  The overall errors of closure phase sets ranged between 6-15 degrees.  In addition, this method of paring off bad data typically reduced errors by a factor of two over other methods.

\footnotetext{The wrapped normal distribution is the probability distribution function of the wrapped variable $\theta \equiv x\mod 2\pi$, given by $p_w(\theta) = \sum p( \theta + 2\pi k)$, where $p$ is the unwrapped probability of the unwrapped variable $x$.  The sum is over integer values of $k$ from $-\infty$ to $\infty$.  The wrapped normal distribution, denoted by $WN$ is, $WN(\theta) \equiv \frac{1}{\sqrt{2\pi\sigma^2}} \sum \exp\large[ \frac{-(\theta-\mu-2\pi k)^2}{2\sigma^2} \large]$, with the same summation limits.}

Even after employing this data paring, some sets of closure phase data contained so much noise that no reliable signal could be discerned.  In this case, the closure phase was removed from the set of eighty-four closure phases further analyzed.  For some targets, up to half of the closure phases triangles were removed.  In these circumstances, the {\em uv}-coverage of the data drop allowed the possibility of model aliasing: i.e., that multiple binary configurations yield similar closure phase sets and each fit the data equally well.  When previous observations of the target were available, we used this information to rule out unlikely fits.  When not, we list all fits to the data.

Finally, non-stochastic errors are typically on the order of one to a few degrees.  This contribution is much smaller than the statistical error, and as such overall best fits of our data changed very little whether or not we attempted to calibrate out this component.

\subsubsection*{Modeling the Binary Fit, and the Calculation of Confidence and Contrast Limits}

For each target, we attempted to fit the observed closure phases with a three-parameter binary model (separation $\rho$, orientation $\theta$, and contrast ratio $r>1$).  The best fitting model is the one which maximized the overall likelihood of the data.  Errors in the parameters are calculated from the curvature of the log-likelihood surface at its maximum. 

The strength of our fits were determined by comparing the increase in log-likelihood, $\Delta\log\mathcal{L}$, between a single star fit and a binary star fit for our real data set as compared to many simulated data sets of single stars.  If the real data set has a much higher value of $\Delta\log\mathcal{L}$ than the simulated data sets, we regard the real data set to be indicative of a binary star.  (For purely Gaussian noise, $\Delta\log\mathcal{L}$ is equivalent to $\Delta\chi^2$.)

We simulate measurements of single stars with identical noise properties and {\em uv}-coverage of the candidate binary target.  The measured closure phases of a single star is the sum of three sources: the intrinsic signal of the target, which is zero for a single star; noise fluctuations from various sources which are described by the standard deviations measured on the target; and a non-stochastic systematic error component, which we assume is negligible compared to the stochastic noise of these targets.  (As a check, we also estimated the typical systematic contribution from the measured signal of eight survey targets whose best fits indicated high likelihood for being single stars.  Including this component to simulate single stars had little effect on the overall confidence measurements.)

From this information, we generate ten thousand mock measurements of single stars.  To each, we fit the three-parameter binary model, record the $\Delta\log\mathcal{L}$, and build a distribution of $\Delta\log\mathcal{L}$ that result from single star observations.  

Comparing the value of $\Delta\log\mathcal{L}$ of the data's fit to the simulated distribution yields the probability of false alarm: the probability that our apparent binary fit is an observation of a single star co-mingled with noise.  The confidence that our target is binary is one minus this false alarm probability. 

To calculate our contrast limits, we first add model binary signals to the simulated single star data.  These mock binary signals span a range of separations, orientations, and contrast ratios.  We fit each, determine the fit confidence, and determine, for a given separation, the highest contrast ratio (i.e. dimmest companion) that would be detected with 99.5\% confidence (false alarm probability of .005).  These calculations are discussed in more detail in the Appendix.

\subsubsection*{Calculation of Bayes' Factors}

As an alternative to confidence measure presented in the previous subsection, our group also applied bayesian methods to calculate the Bayes' Factor of each fit, i.e, the odds by which our data favors binary models.  

Using Bayesian comparison, the binary hypothesis is tested by contrasting two probabilities: that the data set would arise from a binary target observation, and that the set would arise from a single star observation.  Expressed mathematically, this is:

\ba
\lefteqn{\frac{Pr(~\rm{binary}~|~\rm{data}~)}{Pr(~\rm{single}~|~\rm{data}~) } = } \nonumber \\
& & \frac{ Pr( ~\rm{star ~ is ~binary}~ )}{Pr( ~\rm{star ~is ~single}~)} \frac{ Pr(~\rm{data}~ | ~\rm{binary}~ ) }{ Pr( ~\rm{data}~ | ~\rm{single}~ )}.
\ea

The first term on the right hand side is an attribute of the survey population -- it is the ratio of the companion fraction to one minus the companion fraction -- and is independent on the data.  

The second term is the {\em Bayes' Factor}, representing the odds by which the data favors one hypothesis over the other.  These probabilities are marginalized (and integrated) over the binary parameters.  Whereas the maximum likelihood method searches out the set of parameters that maximizes the likelihood of the data, the Bayesian approach {\em averages} over the parameters.

\be
Bayes'~Factor~=  \frac{Pr(~\rm{binary}~|~\rm{data}~)}{Pr(~\rm{single}~|~\rm{data}~) }
\ee
\be
= \frac{ \int Pr( \rho|~{\rm bin.} ) Pr( \theta|~{\rm bin.} ) Pr( r|~{\rm bin.}) \mathcal{L}(\rm{data}|\rho, \theta, r) }{\mathcal{L}( \rm{data}~ |~\rm{single})} 
\ee

The quantities $Pr(\rho|\rm{binary})$,  $Pr(\theta|\rm{binary})$,  $Pr(r|\rm{binary})$ refer to distributions of the companion separation, orientation, and contrast ratio as they are presumed known prior to our observation.  These distributions for very low mass primaries are themselves ongoing topics of debate and limited by observational incompleteness, particularly in the separation regime of our survey.  (For a current review of separation and mass ratio distributions derived through observational studies, the reader is invited to view \citet{Burgasser:2006}.)  \citet{Allen:2007} quantifies the underlying companion distributions from the currently available data.  We ultimately chose to use blind prior distributions (known as Jeffreys' priors).  These distributions are uniform for separation and log-uniform for contrast ratio.  We compare this choice to the Allen priors and discuss its implications.

\begin{figure}[h]
\includegraphics[width=0.45\textwidth]{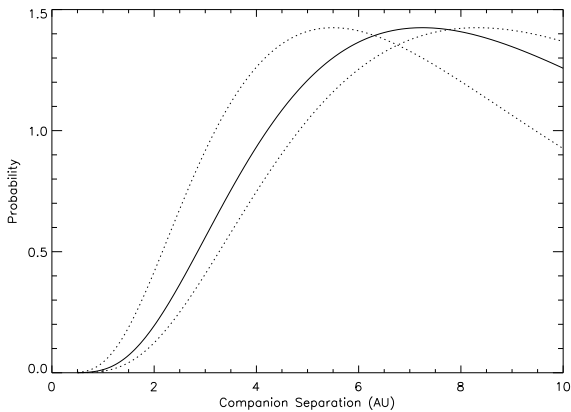}
\caption{Proposed log-normal distribution of companion separation around L dwarf primaries from \citet{ Allen:2007}.  The peak and width of the distribution have been constrained by previous surveys.  The most likely distribution (solid line) and one sigma distributions (dashed lines) are shown.  Despite the constraints, the distribution is noticeably uncertain in the region of separations searched by our survey.  We opt to use a uniform prior for our Bayesian analysis, noting that such a prior may over signify companions closer than roughly 2 AU as compared to the Allen prior.  Similarly, a confirmed detection of a close companion could indicated this distribution has been incorrectly described as log-normal (see text).\label{prior_sep}} 
\end{figure}

Our survey focuses on close binaries; our observations probe between roughly 1 and 8 AU for seventy five percent of our L dwarf targets.  \citet{ Allen:2007} concludes that the (physical) separation of companions can be characterized by a log-normal distribution which peaks at 7.2$^{+1.1}_{-1.7}$ AU with a 1$\sigma$ width of roughly 11$^{+\sim2}_{-\sim3}$ AU.  This uncertainty in the peak and width contributes noticible variability of the resulting distribution at the separations we consider (see Figure \ref{prior_sep}).  One characteristic unifying the span of distributions, however, is that companions closer than $\sim$ 2 AU are less likely by up to an order of magnitude.  The is due in part to describing the distribution as a log-normal functional.  This choice is motivated by the sharp drop in companion fraction observed outward of about 10 AU, and by assuming a similar drop shortward of a few AU, where observational data is incomplete.  As Allen states, this result is derived without well-defined searches for companions at close separations, and the preliminary results of the \citet{Jeffries:2005} and \citet{Basri:2006} surveys potentially indicate the presence of a larger number of close binaries.  We use a uniform prior to avoid the bias of the Allen priors, keeping in mind that companions with high Bayes' factor at less than 2 AU could indicate observational evidence of close companions yet may also be biased towards undue significance.

Observational surveys of very low mass systems show a tendency towards equal mass binaries (q $\sim$ 1) \citep{Burgasser:2006, Reid:2006,  Allen:2007}.  The distribution of mass ratios has been roughly characterized by a power law, $p(q) \propto q^\gamma$, with $\gamma\sim2-4$ depending on the survey.  The large exponent of this distribution indicates that low mass ratio (low q) systems are highly unlikely (and rare).  Transforming this to a distribution of {\em broad-band contrast ratios} ($r$, with $r>1$) requires assumptions about target age distributions, bolometric luminosity corrections, and mass-luminosity models (see \citet{Allen:2005, Allen:2003} for these assumptions applied to field stars).  We wish our prior distribution not depend so highly on these assumptions and rather rely on a few basic assumptions.

The rapid drop of the L dwarf mass-luminosity relation (i.e., halve the mass of the star and its luminosity drops by much more) implies that ratios of contrast are larger than ratios of mass, and that contrast ratios still favor unity (i.e., $p(r) \propto (1/r)^\gamma$ with $0 < \gamma \lesssim 2-4$).  The blind prior for a scale independent quantity like contrast ratios is $p(r) \propto 1/r$ which, conveniently, has the desired properties.  It is worth noting that \citet[Fig. 14]{ Allen:2007} carries out the transformation from mass ratio to contrast ratio, finding a distribution that follows roughly $p(r) \propto 1/(r~\log~r)$ for contrasts down to below 100:1. 

Finally, the same methods can be applied to calculate posterior distributions for $\rho$, $\theta$, and $r$ for each data set.  For a data set with a single best fit, this distribution yields a p.d.f.~describing the best fitting parameters.  The parameter values and errors quoted in this paper are those derived from  maximum fit likelihood, as discussed in the previous subsection, and are not drawn from Bayesian posteriors.  However, we calculate the posterior distributions to assure both methods give comparable results.

\section{ Sixteen Brown Dwarf Targets - Four Candidate Binaries}
\label{results}
\begin{table*}[!thp]
\begin{center}
\caption{The Sixteen Very Low Mass Survey Targets\label{targetlist}}
\footnotesize
\begin{tabular}{ l c c c r @{ $\pm$ } l c c c r c r r c c }
\hline 
\hline
 &R.A. & Decl. & & \multicolumn{2}{c}{Distance}  & J & H & K & \multicolumn{2}{l}{5 Gyr} & \multicolumn{2}{r}{~~~~q~(m$_s$/m$_p$)$^b$} &\multicolumn{2}{r}{1 Gyr} \\
Name & (J2000.0) & (J2000.0) & Spectral Type & \multicolumn{2}{c}{(pc)} & (mag) & (mag) & (mag) & \multicolumn{3}{c}{\tiny 65-105/105-450 mas}& \multicolumn{3}{c}{\tiny 65-105/105-450 mas} \\
\hline
2M 0015+3516......     	&00 15 44.76&	 +35 16 02.6	&L2		&20.7&3.2$^a$&13.88&12.89&12.26&	\multicolumn{3}{c}{0.87~/~0.84}&\multicolumn{3}{c}{0.68~/~0.60}\\
2M 0036+1821$^{s_1}$... 		&00 36 16.17&	 +18 21 10.4	&L3.5	&8.76&0.06&12.47&11.59&11.06&		\multicolumn{3}{c}{0.86~/~0.83}&\multicolumn{3}{c}{0.61~/~0.58}\\
2M 0045+1634$^{s_1}$... 		&00 45 21.43&	 +16 34 44.6	&L3.5	&10.9&2.1$^a$&13.06&12.06&11.37&	\multicolumn{3}{c}{0.83~/~0.82}&\multicolumn{3}{c}{0.57~/~0.54}\\
2M 0046+0715......   		 &00 46 48.41&  +07 15 17.7	&M9		&30.5&4.1$^a$&13.89&13.18&12.55&	\multicolumn{3}{c}{0.86~/~0.83}&\multicolumn{3}{c}{0.80~/~0.73}\\
2M 0131+3801...... 		&01 31 18.38&	 +38 01 55.4	&L4		&20.9&4.2$^a$&14.68&13.70&13.05&	\multicolumn{3}{c}{0.92~/~0.91}&\multicolumn{3}{c}{0.74~/~0.70}\\
2M 0141+1804......		&01 41 03.21&	 +18 04 50.2	&L4.5	&12.6&2.7&13.88&13.03&12.50&		\multicolumn{3}{c}{0.87~/~0.84}&\multicolumn{3}{c}{0.66~/~0.63}\\
2M 0208+2542$^{s_2}$...		&02 08 18.33&	 +25 42 53.3	&L1		&25.3&1.7&13.99&13.11&12.59&		\multicolumn{3}{c}{0.84~/~0.82}&\multicolumn{3}{c}{0.62~/~0.56}\\
2M 0213+4444$^{s_1}$... 		&02 13 28.80&	 +44 44 45.3	&L1.5	&18.7&1.4&13.50&12.76&12.21&		\multicolumn{3}{c}{0.83~/~0.82}&\multicolumn{3}{c}{0.56~/~0.53}\\
2M 0230+2704...... 		&02 30 15.51&	 +27 04 06.1	&L0		&32.5&4.0$^a$&14.29&13.48&12.99&	\multicolumn{3}{c}{0.88~/~0.87}&\multicolumn{3}{c}{0.78~/~0.75}\\
2M 0251-- 0352$^{s_1}$......   		 &02 51 14.90&	 -03 52 45.9	&L3		&12.1&1.1&13.06&12.25&11.66&		\multicolumn{3}{c}{0.92~/~0.91}&\multicolumn{3}{c}{0.75~/~0.71}\\
2M 0314+1603$^{s_1}$...		&03 14 03.44&	 +16 03 05.6	&L0		&14.5&1.8$^a$&12.53&11.82&11.24&	\multicolumn{3}{c}{0.82~/~0.80}&\multicolumn{3}{c}{0.60~/~0.54}\\
2M 0345+2540$^{s_2}$...		&03 45 43.16&	 +25 40 23.3	&L1		&26.9&0.36&14.00&13.21&12.67&		\multicolumn{3}{c}{0.83~/~0.81}&\multicolumn{3}{c}{0.57~/~0.53}\\
2M 0355+1133$^{s_1}$...   	 	&03 55 23.37&	 +11 33 43.7	&L5		&12.6&2.7$^a$&14.05&12.53&11.53&	\multicolumn{3}{c}{0.91~/~0.90}&\multicolumn{3}{c}{0.77~/~0.72}\\
2M 0500+0330$^{s_1}$... 		&05 00 21.00&	 +03 30 50.1	&L4		&13.1&2.6$^a$&13.67&12.68&12.06&	\multicolumn{3}{c}{0.91~/~0.89}&\multicolumn{3}{c}{0.72~/~0.65}\\
2M 2036+1051$^{s_1}$... 		&20 36 03.16&	 +10 51 29.5	&L3		&18.1&3.2$^a$&13.95&13.02&12.45&	\multicolumn{3}{c}{0.87~/~0.85}&\multicolumn{3}{c}{0.63~/~0.58}\\ 
2M 2238+4353...... 		&22 38 07.42&	 +43 53 17.9	&L1.5	&21.8&1.6&13.84&13.05&12.52&		\multicolumn{3}{c}{0.84~/~0.82}&\multicolumn{3}{c}{0.57~/~0.54}\\
\hline
\end{tabular}
\end{center}
\tablecomments{Coordinates and characteristics of the sixteen very low mass targets observed in this sample.  Photometry is taken from the 2MASS catalog.  Spectral types (spectroscopic) and distances are taken from DwarfArchives.org, unless otherwise noted. $^a$Distance measurements derived from J-band photometry and M$_J$/SpT calibration data of \citet{Cruz:2003} assuming a spectral type uncertainty of $\pm$1 subclass. $^b$Survey detection limits of Table \ref{contlimits} given in terms of secondary-primary mass ratio, assuming a co-eval system (same age and metalicity).  Masses ratios are derived from the 5-Gyr (first row) and 1-Gyr (second row), solar-metalicity substellar DUSTY models of \citet{Chabrier:2000}, using J and K band photometry. $^{s_1}$Target previously observed by \citet{Reid:2006}.  $^{s_2}$Target previously observed by \citet{Bouy:2003}}
 \end{table*}
 
 \begin{table*}[!thp]
\centering
\caption{Survey Contrast Limits ($\Delta$K) at 99.5\% Confidence\label{contlimits}}
\begin{tabular}{ l c c c c c c c c c c c c c  }
\hline 
\hline
& \multicolumn{13}{c}{$\Delta$K$^a$} \\
Primary         &  65.0&  85.0& 105.0& 125.0& 145.0& 165.0& 185.0& 225.0& 265.0& 305.0& 345.0& 385.0& 425.0\\
\hline
2M 0015+3516    &  0.92&  1.52&  1.73&  1.88&  2.07&  2.25&  2.27&  2.18&  2.03&  1.82&  2.06&  2.06&  1.79\\
2M 0036+1821    &  1.77&  2.30&  2.52&  2.57&  2.63&  2.74&  2.77&  2.79&  2.71&  2.70&  2.62&  2.67&  2.56\\
2M 0045+1634    &  2.06&  2.61&  2.82&  2.87&  2.90&  2.96&  3.01&  3.02&  2.94&  2.93&  2.80&  2.86&  2.84\\
2M 0046+0715    &  0.62&  1.01&  1.16&  1.29&  1.48&  1.58&  1.66&  1.60&  1.38&  1.27&  1.28&  1.35&  1.22\\
2M 0131+3801    &  0.74&  1.26&  1.30&  1.30&  1.35&  1.47&  1.52&  1.55&  1.45&  1.28&  1.25&  1.41&  1.25\\
2M 0141+1804    &  1.52&  2.13&  2.37&  2.51&  2.55&  2.61&  2.65&  2.59&  2.58&  2.58&  2.41&  2.51&  2.42\\
2M 0208+2542    &  1.29&  1.93&  2.16&  2.28&  2.35&  2.48&  2.52&  2.47&  2.34&  2.28&  2.29&  2.32&  2.25\\
2M 0213+4444    &  1.84&  2.40&  2.59&  2.64&  2.72&  2.77&  2.79&  2.81&  2.76&  2.73&  2.59&  2.69&  2.58\\
2M 0230+2704    &  0.72&  1.11&  1.20&  1.18&  1.23&  1.27&  1.30&  1.32&  1.27&  1.09&  1.03&  1.26&  1.08\\
2M 0251-0352     &  0.69&  1.07&  1.24&  1.26&  1.32&  1.39&  1.48&  1.38&  1.36&  1.28&  1.24&  1.37&  1.34\\
2M 0314+1603    &  1.52&  2.08&  2.32&  2.47&  2.54&  2.60&  2.65&  2.61&  2.54&  2.50&  2.51&  2.52&  2.39\\
2M 0345+2540    &  1.75&  2.28&  2.51&  2.56&  2.61&  2.70&  2.76&  2.75&  2.61&  2.57&  2.55&  2.58&  2.51\\
2M 0355+1133    &  0.69&  1.15&  1.21&  1.05&  1.04&  1.22&  1.27&  1.25&  1.29&  1.32&  1.24&  1.27&  1.12\\
2M 0500+0330    &  0.79&  1.35&  1.53&  1.64&  1.81&  1.97&  2.02&  1.99&  1.80&  1.57&  1.80&  1.83&  1.57\\
2M 2036+1051    &  1.30&  1.90&  2.10&  2.26&  2.31&  2.41&  2.50&  2.40&  2.31&  2.24&  2.26&  2.30&  2.18\\
2M 2238+4353    &  1.77&  2.29&  2.51&  2.53&  2.57&  2.63&  2.71&  2.69&  2.57&  2.54&  2.52&  2.56&  2.50\\
\hline
\end{tabular}
\tablecomments{Detection contrast limits around primaries: $^a$Primary-Secondary separations are given in units of mas, and the corresponding detection limits are in $\Delta$K magnitudes.}
\end{table*}

 \begin{table*}[!thp]
\centering
\caption{Model Fits to Candidate Binaries\label{binaries}}
\begin{tabular}{l c r @{ $\pm$ } r r @{ $\pm$ } l r @{ $\pm$ } l c c r @{ $\pm$ } l }
\hline
\hline
               & J. Date & \multicolumn{2}{c}{Separation} & \multicolumn{2}{c}{Az. Ang.} & \multicolumn{2}{c}{Contrast} &  Bayes &  &  \multicolumn{2}{c}{Separation} \\
Primary &(+245000) & \multicolumn{2}{c}{(mas)} & \multicolumn{2}{c}{(deg)} & \multicolumn{2}{c}{Ratio}  & Factor & Conf.&  \multicolumn{2}{c}{(AU)} \\
\hline
2M 0036+1821      & 4731 & 89.5   & 11.4 & 114.1 & 5.5 & 13.14 & 3.14 & 7.9 & 96\% & 0.78 & 0.10\\ 
\hline
\multirow{2}{*}{2M 0345+2540}      & \multirow{2}{*}{4731} & 217.4 & 9.1 & 258.8 & 2.8  & 26.44 & 4.22 & \multirow{2}{*}{7.6}   & 98\% & 5.85 & 0.26 \\
                            				      &        			           & 352.7 & 10.5 & 87.6 & 2.0 & 30.79 & 9.08 &         			         & 96\%  & 9.49 & 0.31\\
\hline
\multirow{3}{*}{2M 2238+4353}      & \multirow{3}{*}{4732} & 128.2 & 10.3 & 209.9 & 5.3 & 17.76 & 4.25 &  \multirow{3}{*}{7.1} & 97\% & 2.79 & 0.30 \\
                    				      &         				  & 228.5 & 9.1 & 251.8 & 3.5   & 23.79 & 5.92 &      			         & 95\% & 4.98 & 0.42 \\
                   			               &           			  & 395.5 & 9.7 & 19.5   & 1.2   & 17.63 & 4.22 &    			         & 97\%  & 8.62 & 0.66\\
\hline
2M 0355+1133      & 4757 & 82.5 & 13.0 & 276.2 & 4.1  & 2.10   & 0.40 &  6.3  & 90\% & 1.03 & 0.27\\
\hline
\end{tabular}
\end{table*}
		
\begin{figure*}[!thp]
\includegraphics[width=\textwidth, height=5.5cm]{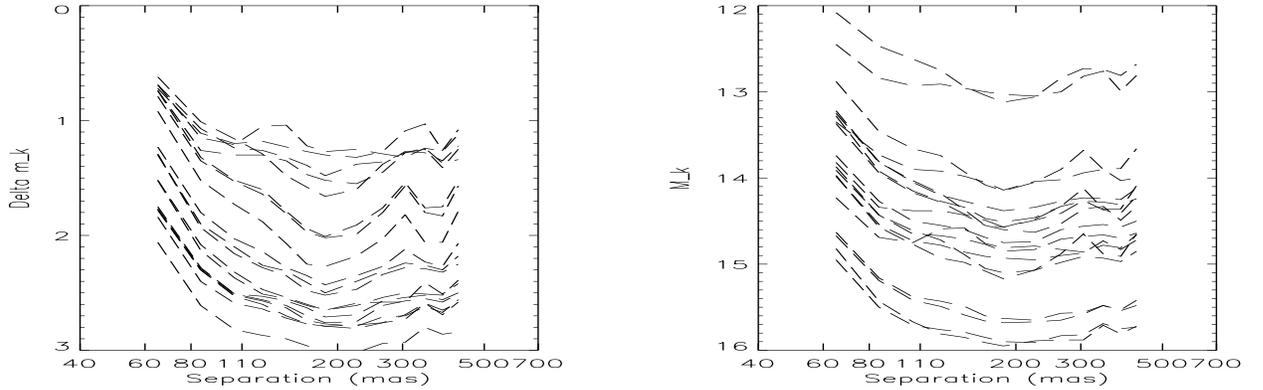}
\caption{Contrast limits at 99.5\% detection as a function of primary-companion separation: (left) The primary-secondary magnitude difference in K$_s$ detectable at 99.5\% confidence.  (right) The same detection limits in terms of the absolute magnitude of the companion.\label{contlimits_fig}} 
\end{figure*}

\begin{figure*}[!thp]
\plotone{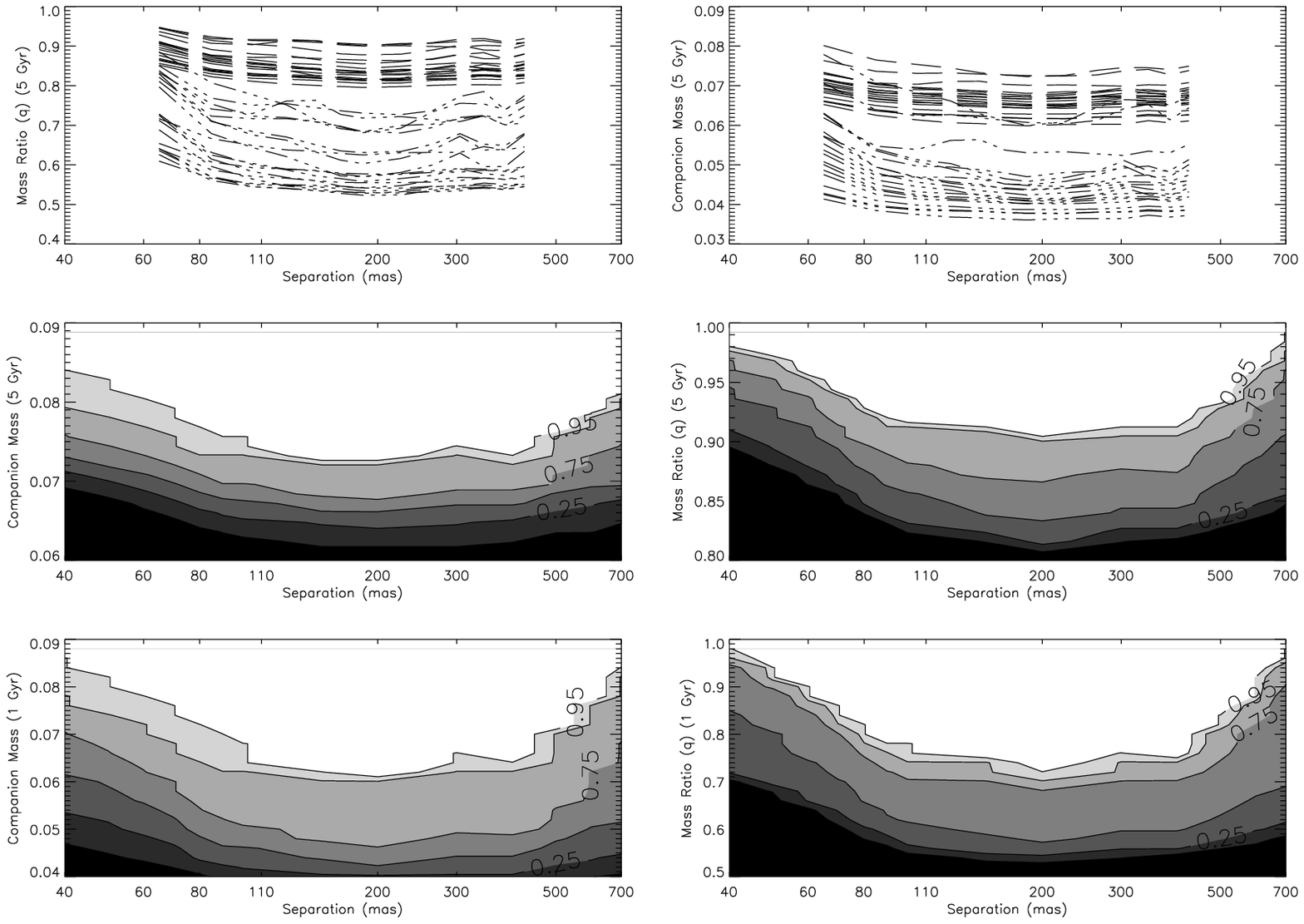}
\caption{Companion mass and mass ratio limits at 99.5\% detection as a function of primary-companion separation: (top left) The primary-companion mass ratio detectable at 99.5\% confidence.  {\em Dashed} lines are for systems aged 5 Gyr; {\em Dot-dashed} lines are systems ages 1 Gyr.  (top right) The same data in terms of companion mass.  (middle/bottom left) As a function of separation and companion mass, this plot reveals the percentage of 5 Gyr (middle) and 1 Gyr (bottom) companions detectable at 99.5\% given the data quality of the survey.  Binaries in the white area would have been detected for 100\% of the survey targets, followed by contour bands of 95\%, 90\%, 75\%, 50\%, 25\%, and 10\%.  At the diffraction limit (110 mas), companions of mass 0.65 \msun would be resolved for 50\% of our targets.  (middle/bottom right)  The same plot in terms of mass ratio.   Diffraction limit sensitivity: 5 Gyr companions of mass 0.65 \msun (.038 \msun for 1 Gyr) would be resolved for 50\% of our targets.  Equivalently, our survey reached mass ratios of .83 (5 Gyr) and .55 (1 Gyr) for 50\% of our targets at the diffraction limit.\label{mass_fig}} 
\end{figure*}

	Aperture masking is most sensitive to companions between $\lambda / 2D$ and $4\lambda / D$, corresponding to angular separations of 60-450 mas in K$_s$ at Palomar and physical projected separations ranging from 0.6-10 AU for the targets in our survey.  
	
	Our achieved detection limits for all sixteen targets are summarized in Table \ref{contlimits}.  Our limits remain relatively flat at separations beyond $\lambda/D$, plateauing near $\Delta$K$\sim$2.3 for more than half our targets, and decline to roughly 1.4 magnitude shortward of $\lambda/D$ (See figure \ref{contlimits_fig}).
	
	We infer the (companion) stellar properties and mass ratios to the corresponding magnitude limits using the DUSTY models for target ages of 5 Gyr and 1 Gyr (Table \ref{targetlist}).  At the formal diffraction limit (about 110 mas in K$_s$), companions with mass ratios of .83 for 5 Gyr systems and .55 for 1 Gyr systems would be resolved for 50\% of our targets at a 99.5\% confidence of detection (Fig. \ref{mass_fig}).
	
	Our survey found four candidate binary systems with detections at 90-99\% confidence and Bayes' Factors favoring the binary model (Table \ref{binaries}).  We summarize and discuss these detections below.
	
	 For some targets in our survey, closure phase measurements constructed from the longest baselines had too much noise to extract a useable signal.   The resulting drop in {\em uv}-coverage can give rise to aliasing of the model fits: i.e., multiple binary configurations fit the data equally well.  When possible, we used previous observations of the target to rule out certain aliased fits; when not possible, all model fits are listed.

	{\textbf{2M 0036+1821:}}  A companion at separation 89.5 mas and 13.1:1 contrast was detected with 96\% confidence and a Bayes' Factor of 7.8:1.  The data also fits an alternative (alias) binary configuration ($\rho \sim$ 243 mas and 25:1 contrast) with 96\% confidence that we rule out by a previous observations.  \citet{Reid:2006} observed this target in November 2005 with the NICMOS imager on the Hubble Space Telescope in the F170M and F110W bands.  At or near this separation, this alternative configuration would have likely been detectable in the F110W bands.
	
	{\textbf{2M 0355+1133:}}  A companion at separation 82.5 mas and 2.1:1 contrast was detected with 90\% confidence and a Bayes' Factor of 6.3:1.  \citet{Reid:2006} also observed this target in the F110W band and found no companion.  As a proxy for the F110W bandpass, we estimate a J band contrast of 2.5:1 using the J-K color-magnitude relations of \citet{Dahn:2002}.  Their program achieved a contrast limit of 2.5:1 beyond approximately 100 mas in F110W, suggesting that this candidate binary sat at the edge of their detection limits.
	
	{\textbf{2M 2238+4353:}}  Thirty-five percent of the closure phase triangles showed very high noise and were removed from analysis.  As a result, aliasing of the signal was particularly problematic.  Three distinct binary configurations were detected at 95-97\% confidence.  These range in separations between 100 and 400 mas and contrasts between 17:1 and 28:1.  
	
	{\textbf{2M 0345+2540:}} Like 2M 2238+4353, a large percentage of the closure phases were removed from analysis.  Two distinct configurations, both with contrasts $\sim$28:1 ($\Delta$K $\sim$ 3.5) were determined with high confidence. \citet{Bouy:2003} observed this target with the Wide Field Planetary Camera 2 (WFPC2) on the Hubble Space Telescope in March 2001, but we estimate these companions to be below their detection limits.  Their survey reached background limitations at contrasts between $\Delta M \sim$3-5 in the F814W band.  Using the I band as a proxy for F814W, we estimate the companion of 2M 0345+2540 to have a contrast of $\Delta I \gtrsim$5 and to have been undetectable in the Bouy survey. 
	
	{\textbf{2M0213+4444:}} We observed target 2M0213+4444 three times over two nights in September 2008 (two sets in K$_s$, one in H) and once one month later (in K$_s$).  Two data sets from September were of poor quality and were not used for analysis.  The remaining set from September found one binary fit ($\rho\sim$ 81 mas, $\theta\sim$ 290$^\circ$, 5.2:1 contrast in K$_s$) at 89\% confidence.  The target was observed again in $K_s$ in October under poor seeing and much of the data was unusable.  This data set could not be fit well by the September results, and implied a different configuration with 90\% confidence ($\rho\sim$ 234 mas, $\theta\sim$ 135$^\circ$, 11:1 contrast in K$_s$).  Given the low confidences of fits and the unreproducibility of these results, we conclude that this target is unlikely to be binary.

\section{Discussion: Aperture Masking of Faint Targets}
\label{discussion}

	The use of non-redundant masking removes many types of spatial perturbations to the incoming wavefront.  During high signal to noise observations, when read and background noise are minimal, the largest contributor to measurement noise is the temporal and spatial atmospheric fluctuations of the wavefront, even after adaptive optics correction.  Short exposure times, roughly less than the coherence time of atmosphere, freezes the tip-tilt and low-order perturbations to the wavefront, which can be removed by combining fringe phases into closure phases.  This advantage is lost when exposures extend over multiple coherence times.  For this reason, aperture masking flourishes with short exposures.  
	
	Behind laser guide star adaptive optics systems, although the structure of the corrected wavefront may be different, the functionality of aperture masking is the same.  However, targets requiring laser guide star AO tend to be fainter, and thus require either longer exposure times (permitting sufficient correction) or techniques for dealing directly with noise from read outs and background flux.
	
	This survey opted for maintaining short exposure times.  The signal to noise of fringe amplitudes decline rapidly for longer baselines, as the transmission function for these baselines is lower and turbulence variations are larger.  Just as, for instance, Stehl ratio depends on the variance of the incoming wavefront, so does the fringe amplitude, also dropping as $\exp( -\sigma_{baseline}^2 )$.  For faint targets, long baselines fringes often linger undetectable below the background and read noise, making difficult measurements of long baseline phases. 

	The capture of a large number of short exposure images allows us to select out the best fringe measurements, during the serenditous moments of very good correction or still atmospheres, and discard those dwarfed by read noise.  This technique, analogous to {\em lucky imaging}, effectively selects high signal to noise measurements of closure phase.  In most cases, these lucky closure phases were sufficient to obtain measurements of the target closure phase, even at long baselines.  	
	
	We contrast this method to two measurements of targets observed with long (1 minute) exposures.  These exposures did often have long baseline fringes detectable at or just above background.  But this method resulted in poor measurements of the target closure phase, even at shorter baselines.  The multiple-coherence time exposures means that low order perturbations are not effectively removed by closure phases, resulting in large phase errors, and the fewer overall number of data points removes the statistical advantage.  The measured closure phase is not a good measurement of the true target phase.
	
	Long exposures, with adequate correction, do allow longer baseline fringes to grow in amplitude above the read noise or background limit.  Exposures for aperture masking should be limited to the effective coherence time of the adaptive optics system -- the interval over which the phase variance of the longest baselines reaches about one radian.  
	
	The quality of measurements from both sets of exposure data suggests a slight modification of technique for the next application of aperture masking with laser guide star adaptive optics.  The higher noise content of the one-minute exposures suggests that these exposures are too long for the level of correction obtained in this survey.  The short exposure method fared much better, but a large percentage of images failed to observe fringe amplitudes above read noise.  This suggests that slightly longer exposures would have benefited the observations.  It is worthwhile to note that the low transmission of the long baselines, even at Strehl ratios of 15\% typically reached in this survey, indicates that direct imaging would not have been able to obtain $\lambda/D$ resolution.	

\section{Conclusion}
\label{conclusion}

	We present the results of a close companion search around nearby L dwarfs using aperture masking interferometry and Palomar laser guide star adaptive optics.  The combination of these techniques yielded typical detection limits of $\Delta$K$_s$ = 1.5-2.5 between 1-4$\lambda/D$ and limits of $\Delta$K$_s$ = 1.0-1.7 at 0.6 $\lambda/D$.  Our survey revealed four candidate binaries with moderate to high confidence (90-99\%) and favorable Bayes' Factors.
	
	Ten of the targets have previously been observed with the Hubble Space Telescope as part of companion searches.  As such, we did not expect to find bright or distant companions around these targets which would have been identified in the previous surveys.  But as demonstrated in this paper, the detection profile of aperture masking is capable of revealing close or dim companions which are obscured by the point spread function of full aperture imaging.  Aperture masking demonstrates an increase in formal resolution and detectable contrast at close separations over laser guide star adaptive optics alone.
		
	Our survey indicated two previously observed targets as candidate binaries.  Our survey indicated one companion around 2M 0355+1133 within the formal diffraction limit of the HST and one companion around 2M 0345+2540 below the background detection threshold of the previous survey.  Two other targets, 2M 0345+2540 and 2M 2238+4354, also indicated the presence of companions, both with contrast ratios greater than 15:1.  
	
	Aperture masking is most sensitive to companions between $\lambda / 2D$ and $4\lambda / D$, corresponding to angular separations of 60-450 mas in K$_s$ at Palomar and physical projected separations ranging from 0.6-10 AU for the targets in our survey.  Two candidate binaries presented in this paper have projected separations less than 1.5 AU.  The results suggest a favorable target set for future companion searches.  Their candidacy is consistent with the conjecture that tight binaries are underrepresented in the current tally of low mass binaries.  Spectroscopic surveys, which focus on separations within 3 AU, are necessary to conclusively answer this question.  Extending the use of aperture masking with laser guide star AO is a rewarding approach for detecting companions within this range, and facilitating the measurements of their masses.

\acknowledgements

We thank the staff and telescope operators of Palomar Observatory for their support.  David Bernat thanks Jason Wright for many helpful discussions about the analytical techniques developed within this paper.  We thank the Palomar staff for many nights of assistance at the Palomar Hale telescope.  This work was supported in part by the National Science Foundation under award numbers AST-0905932 and AST-0705085.  The Hale Telescope at Palomar Observatory is operated as part of a collaborative agreement between Caltech, JPL, and Cornell University.  This publication makes use of data products from the Two Micron All Sky Survey, which is a joint project of the University of Massachusetts and the Infrared Processing and Analysis Center/California Institute of Technology, and is funded by the National Aeronautics and Space Administration and the National Science Foundation.  

Facilities: \facility{Palomar(LGSAO)}

\appendix
\label{appendix}

\section{Binary Detection Confidence and Contrast Limits}

In this section, we provide the basic formalism for fitting closure phase measurements to binary models.  This formalism also  drives the Monte Carlo simulation which determines the strength of these fits.   

Each pair of holes in the Palomar non-redundant nine-hole aperture mask is designed to transmit one unique Fourier spatial frequency, for a total of thirty-six frequencies transmitted by the mask.  Each image produced by the mask consists of thirty-six overlapping fringes which, when Fourier transformed yields the amplitude and phase of each transmitted frequency.  Combining these phases by adding specific triplets to form closure phases produces an observable that is more robust to many forms wavefront noise.  Eighty-four closure phases are extracted from each image, then averaged over the set of images.  Finally,  these averages are compared to model closure phases of various binary configurations to determine the likelihood that the target is binary.

\subsection*{Determination of Best Fit}

The measured closure phase signal of each image is the composite of three sources: the intrinsic signal of the target, which is zero for a single star and non-zero for a binary; a non-stochastic systematic error component, which may vary from target to target, but not during the observation of a single target; and stochastic noise from various sources such as atmospheric turbulence, read noise, etc.  We denote the intrinsic signal by $\Psi_{binary}$, the systematic component by $\beta_{system(t)}$, and the stochastic noise by $\xi_{noise(t,i)}$.  That is, the measurement of closure phase $k$, extracted from image $i$, observed during the target acquisition set $t$, $\Psi_{k,t,i}$, is:

\be
\Psi_{k,t,i} = \Psi_{k,binary} + \beta_{k,system(t)} + \xi_{k,noise(i,t)}. \\
\ee

The measured closure phase from a given set of images that will be fit to model binaries is 
\be
< \Psi_{k,t} >_i = \Psi_{k,binary} + \beta_{k,system(t)},
\ee
where the sum is taken over the set of images, and the noise properties of the mean is described the by distribution of $\xi_{k,noise(i)}$ convolved over the set of images.  The systematic component $\beta_{k,system(t)}$ may change from one acquisition to another (affected by telescope movement or AO system performance, etc.) but is assumed constant during the observation of a single target.

We typically use the measurement of calibrator (single) stars, with zero intrinsic signal (i.e., $\Psi_{k,binary}=0$), to estimate the underlying distribution of systematic noise in the optical system.  The typical observing mode is to obtain several observations of the science target, interspersed with observations of calibrator stars.  Although the systematic component cannot be determined exactly because it is itself a random variable, we can compile a composite distribution of $\beta_{k,system}$ from calibrator observations.  Subtracting the systematic component from the measured closure phases (or, specifically, convolving the two distributions), leaves remaining the intrinsic signal of the target, $\Psi_{k,binary}$.

(This calibration step is important for obtaining high contrasts during high signal to noise observations, when the contribution from systematic noise is on the order of the stochastic noise.  Calibration is less effective when the typical stochastic noise is much larger the systematic component.)

We wish find the three-parameter model binary (separation $\rho$, orientation $\theta$, and contrast ratio $r$) which best fits the calibrated signal, $\Psi_{k,binary}$.  Approached as a maximum-likelihood problem, we calculate the probability that our measured data would result from a noiseless, modeled binary, $\Psi_{m}(\rho, \theta, r)$, observed through the noise in the system:

\be
\mathcal{L}_{model} = p( \Psi_m | \{\Psi_{k,binary}\} ) \propto \prod_k \int \mathcal{L}_k( \Psi_{k,t} = \Psi_m - \beta_{k,system}) ~p( \beta_{k,system} )~d\beta_{k,system} \label{pmodel}
\ee

The integral is due to the convolution of the systematic distribution with the likelihood function, which itself follows the distribution of $\xi_{k,noise}$.

The best-fitting model is that which maximizes the above probability, which we determine by a combination of gradient search and visual inspection.  The parameter errors are calculated from the curvature of the log-probability surface at the maximum.  Calculating the confidence of this fit, i.e. that this model represents the true target configuration, is detailed in the next subsection.

Assuming the underlying distributions are wrapped-normal or Gaussian, the convolution above reduces to a single wrapped-normal or Gaussian distribution with mean $<\Psi_{k,t}> - <\beta_{k,system}>$ and variance $\sigma^2_{\Psi_{k,t}}+\sigma^2_{\beta_{k,system}}$.  The maximum probability problem reduces to one of minimizing $\chi^2$.

\subsection*{ Binary Detection Confidence}

Our null hypothesis, which we wish to test against the binary fit, is that the observed target is a single star, with intrinsic binary phase zero.  Following Eq  \eqref{pmodel}, the probability of the null model is

\be
\mathcal{L}_{null} = p_{null}( \Psi_m=0 | \{\Psi_{k,binary}\} ) \propto \prod_k \int p( \Psi_{k,t} = \beta_k | \beta_k )~p( \beta_k )~d\beta_k.
\ee

A natural goodness-of-test statistics for comparison is the ratio of the data likelihood to that of the null likelihood, or the log of this ratio:

\be
\log( \frac{\mathcal{L}_{model}}{\mathcal{L}_{null}}) = \log(  \mathcal{L}_{model} ) - \log(  \mathcal{L}_{null} ) = \Delta \log \mathcal{L}.
\ee

Systematic and stochastic noise may at times conspire to mimic a binary signal, as expressed by a higher probability of a binary model, even though the target is a single star.  This is a false alarm event.  We, therefore, classify the target goodness-of-fit statistic as statistically significant only if its value is large compared results of noisy observations of single stars.

We simulate ten thousand measurements of single stars with identical ($u$,$v$)-coverage and noise properties of the candidate binary target.  The intrinsic phase of a single star, $\Psi_{binary}$, is zero.  For one measurement, the contribution due to statistical noise is drawn from the measured distribution of $\xi_{k,noise}$, which typically can be approximated by a wrapped normal or Gaussian distribution with mean zero and measured standard deviation.  The systematic contribution, if included, is drawn from a distribution $\beta_{k,system}$, compiled from observations of calibrator (single) stars.

We then apply the same approach used for the target star to each simulated single star.  We fit the three-parameter binary model to each simulated single star, record the $\Delta\log \mathcal{L}$, and build a distribution of $\Delta\log(\mathcal{L})$ that results from fitting single stars.  The probability that our target fit is statistically significant, then, is the probability that the goodness-of-fit of a single star is less than the target data's goodness-of-fit:

\be
p_{false~alarm}( \Psi_m) = p(~(\Delta\log \mathcal{L})_{best~fit~to~data} < (\Delta\log \mathcal{L})_{fits~to~single~stars}~)
\ee
and
\be
detection~confidence = 1 - p_{false~alarm}.
\ee

\begin{figure}[!h]
\centering
\includegraphics[height=8cm,width=14cm]{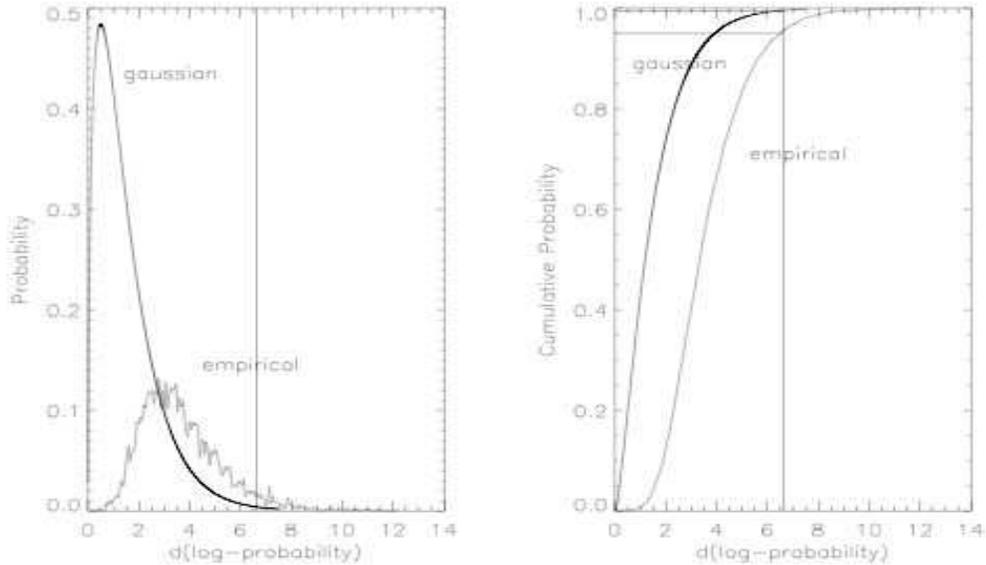}
\caption{Determination of Binary Confidence for Target 2M 0036+1806.  The goodness-of-fit statistic, $\Delta\log\mathcal{L}$ = 6.55 (vertical line), is compared to a distribution generated from fits to simulated single stars, resulting in a fit confidence of 96\% (empirical).  Notice that comparing this value to a $\chi^2$ distribution with three degrees of freedom (Gaussian) results in a much higher confidence of fit.\label{conflim_plot}}
\end{figure}

We consider the target data to reveal a definitive binary detection if the best-fitting model produces a detection confidence greater than 99.5\% (false alarm probability less than 0.5\%).  Note that this empirical method is more conservative than comparing the values of $\Delta\log \mathcal{L}$, which reduces to $\Delta \chi^2$ for Gaussian noise, to a distribution of $\chi^2$ variables with three degrees of freedom (Fig. \ref{conflim_plot}).

\subsection*{Calculation of Contrast Limits}
 
Whether the target observed is identified as a single star or binary, we are able to quantitatively state the highest contrast (dimmest) companion that our technique {\em would have} been capable of identifying with high confidence (99.5\%) as a function of separation.  This is, in essence, a statement on the noise characteristics of the data and the {\em uv}-coverage of our mask.

This amounts to asking the following question:  Given simulated binary observations (separation $\rho$, orientation $\theta$, and contrast ratio $r$), at what contrast does our detection confidence drop below 99.5\%?

Simulated binary data is the composite of the same noise contributions to single star data plus an intrinsic signal due to the presence of a companion.  That is, the $n$th simulated binary data is: 

\be
\Psi_{k,binary}^n = \Psi_{k,single}^n + \Psi_{k,model}(\rho, \theta, r)
\ee

For binary models of a range of separations, orientations, and contrast ratios, we generate ten thousand mock binary signals of each by adding the intrinsic binary signal to the mock noise simulations described in the previous subsection.  We fit each, record the fit confidence, and determine the average confidence that that binary can be detected under conditions identical to the target observation.  For each separation, averaged over orientations, we determine the highest contrast ratio (i.e. dimmest companion) that would be detected with 99.5\% confidence (false alarm probability of .005).

Ideally, we would determine the confidence of each mock binary by search for its best fit, recording its $\Delta\log\mathcal{L}$, and comparing it to the false alarm distribution of the previous subsection.  In practice, using a fitting routine to fit each of these simulated binaries is computationally slow.

Instead we approximate this process by modifying the false alarm distribution.  We make the assumption that the inserted binary model yields the best fit.  Because we effectively restrict the fitting search to the range of separation, orientations, and contrast ratios used to generate the mock binaries, we apply the same restriction to the fitting search that generates the false alarm distribution.  We then use this modified false alarm distribution to determine the confidence of the mock binary fits.  In practice, this approximation produces contrast limits slightly more conservative than full fitting by about 5-10\%.

\end{document}